\documentclass[12pt]{article}
\usepackage{amsfonts}
\def\be{\begin{equation}}
\def\ee{\end{equation}}
\def\bea{\begin{eqnarray*}}
\def\eea{\end{eqnarray*}}

\def\pd{\partial}
\def\t{\tau}

\def\w{\omega}
\def\S{\Sigma((0,1])}

\def\integer{\mathbb{Z}} 
\def\real{\mathbb{R}}
\def\natural{\mathbb{N}}
\def\finis{~\rule{.5\baselineskip}{.55\baselineskip}}
 
\newtheorem{lem}{\bf Lemma}[section] 
 
\newtheorem{thm}[lem]{Theorem} 
 
\newtheorem{rem}[lem]{Remark}

\newenvironment{proof}{\noindent\emph{Proof.}}{\hspace*{\fill}{}\bigskip\par}
             
\begin{document} 
\begin{titlepage}
\vspace*{-2cm}
\begin{flushright}
DTP--05/35\\
ANL-HEP-PR-05-71\\

\end{flushright}

\vspace{0.3cm}

\begin{center}
\Large{{Lie algebras associated with one-dimensional \\aperiodic point sets}}

 \vskip 0.6 true cm
{\small
{ David B. Fairlie$^1$,\ \ Reidun Twarock$^2$\ \ and\ \  Cosmas  K. Zachos$^3$}
\vskip 0.5 true cm
{\small
\centerline{David.Fairlie@durham.ac.uk, rt507@york.ac.uk, zachos@anl.gov}}
\vskip 0.2 true cm
{\it $^1$ Department of Mathematical Sciences, University of Durham}\\
{\it South Road, DH1 3LE, Durham, England}\\[3pt]
{\it $2$ Department of Mathematics, University of York}\\
{\it Heslington York YO1 5DD, England}\\[3pt]
{\it $^3$  High Energy Physics Division, Argonne National Laboratory}\\
{\it Argonne, IL 60439-4815, USA}}
\vskip 1.9 true cm 
\noindent 
{\bf Abstract} 
{\small\begin{flushleft} 
The  set of points of a one-dimensional cut-and-project quasicrystal or model set, while 
not additive, is shown to be multiplicative for appropriate choices of 
acceptance windows. This leads to the definition of an associative additive graded composition
law and permits the introduction of Lie algebras over such aperiodic point sets. These infinite dimensional Lie algebras are shown to be representatives of a new type of semi-direct product induced Lie algebras. 
\end{flushleft}}
\end{center}
\vfill
\end{titlepage}

\section{Introduction}

In the study of aperiodic structures, the {\it Fibonacci chain} 
$\mathcal{F}$, a one-dimensional aperiodic sequence,  plays an important 
role. It may be co-ordinatized by the  sequence  
$\{\mathcal{F}: ~\dots, -4\t-2, -3\t-1, -2\t-1,-\t,~1,~\t + 1,2\t + 2, 
3\t+2,4\t+3, 5\t+4,6\t+4,7\t+5,8\t+5,9\t+6,10\t+7, \dots \}$,
which is given for all positive and negative integers $n_2\in \integer$ by 
\be
\mathcal{F}(n_2) = n_2\t + \left\lfloor\frac{n_2}{\t} +1\right\rfloor .
\label{floor}
\ee

The floor function, denoted by $\lfloor x \rfloor$, is the greatest integer
less than, or equal to $x$; whilst 
$\tau=\frac12 (1+\sqrt{5})\approx 1.618$ denotes the {\it golden mean}, 
which satisfies the equation
\be
x^2\,=\,x+1.\label{def}
\ee
It corresponds to the sequence featured in, e.g., 
\cite{senechal,patera,twarock}.

The Fibonacci chain $\mathcal{F}$ in (\ref{floor}) is a subset of the ring $\integer [\tau] := \integer + \integer \tau$, equipped with the Galois automorphism that maps $n=n_1+ n_2 \tau \in \integer [\tau]$ to $n^* = n_1+ n_2 \tau'$, 
where $\tau'\,\equiv \,-\tau^{-1}$ is the algebraic conjugate\footnote{It is the other root of the defining quadratic 
equation (\ref{def}), and so the multiplication table of functions of 
$\tau'$ is identical to that of functions of $\tau$.} of $\tau$. 
$\mathcal{F}$ is an example of a {\it cut-and-project quasicrystal} or {\it model set}, 
i.e., a set of the form 
\be
\Sigma(\Omega)=\{ n\in \integer [\tau] ~~| ~~ n^* \in \Omega \} ,  \label{dual}
\ee
and $\Omega\subset \real$ is a connected interval, called an {\it acceptance window}. 
In particular, the Fibonacci chain $\mathcal{F}$ amounts to $\Sigma((0,1])$. 
While (\ref{floor}) is useful for explicit evaluation of the points in 
$\mathcal{F}$, set theoretic properties of the chain can 
be discussed more conveniently in terms of (\ref{dual}). 

Recently, a new class of infinite dimensional Lie algebras was introduced 
\cite{FZ}, which possess indices valued in cyclotomic fields 
$\integer(\w)$ with $\w^N=1$ an $N$-th root of unity. 
For $n$, $m\in \integer(\w)$ and $a$, $b\in\lbrace 0, \ldots, N-1\rbrace$, 
these algebras are defined by the relation 
\be 
[J^a_{m},J^b_n]\,=\,J^{a+b}_{m+w^an}\,-\,J^{a+b}_{n+w^bm}\,, \label{algs} 
\ee
and may be constructed from the more succinct associative product law
\be J^a_{m}  J^b_n\, =\,J^{a+b}_{m+w^an}.\label{mult}
\ee
This is an algebra in which one of the indices adds with a phase factor, 
unlike the conventional indices of Kac-Moody algebras.

For $N\geq 5$ the set $\integer(\w)$ is dense in $\real^2$, and so is any 
one-dimensional subset. We consider here such one-dimensional subsets, and introduce a condition to select appropriate uniformly discrete subsets of these sets. 
In particular, since $\tau = -(\omega^2 + \omega^3)$ for $\omega^5=1$, the 
sets $\Sigma(\Omega)$ in (\ref{dual}) are, in fact, uniformly discrete 
one-dimensional subsets of $\integer(\w)$ that are located on the real line 
in the complex plane. 

A crucial feature of the sets $\Sigma(\Omega)$ is their lack of additivity,
which does not encourage trivial applications. 
However, it is shown here that, for appropriate choices of $\Omega$, the set 
is {\em multiplicative}. Moreover, a further, associative, 
graded additive composition closure 
holds: For all $m$, $n\in \Sigma(\Omega)$ there exists $a\in \natural$  such that 
$m+\tau^a n \in 
{\Sigma(\Omega)}$.  
In particular, multiplicativity allows one to restrict (\ref{algs}) to a 
subalgebra (Lie algebra) (\ref{mult}), with generators in a one-to-one 
correspondence with points in a one-dimensional aperiodic point set. 

Note that these Lie algebras are different from the known family of aperiodic 
Virasoro algebras constructed over $\Sigma(\Omega)$, proposed in 
\cite{patera,twarock}. Specifically, in these references, an aperiodic analog 
to the Virasoro algebra has been defined via a replacement of  the usual 
index set $\integer$ by a cut-and-project quasicrystal $\Sigma([a,b])$ with 
$a\cdot b \geq 0$. The corresponding Lie algebras are given by the 
relation 
\be
\label{AW} 
[L_n,L_m]= (m-n)\chi_{[a,b]}(n'+m')L_{n+m}\,,
\ee
where $\chi_{[a,b]}$ denotes the characteristic function of the 
set $[a,b]$ \cite{patera}. It has been shown that a nontrivial central 
extension exists for the case of $[a,b]=[0,1]$, if the structure constants 
are restricted to the $\tau$-component of the points $(n_1 +\tau n_2) 
\in \Sigma([0,1])$, i.e. to $n_2$ \cite{twarock}. 
Note that these algebras are by construction nilpotent, due to the factor 
$\chi_{[a,b]}$. By contrast, the algebras defined in this paper are not 
nilpotent.  

\section{Algebras reliant on multiplicative closure of $\Sigma(\Omega)$}

This section relies on the following theorem: 
\begin{thm}\label{thm1}
$\Sigma(\Omega)$ in (\ref{dual}) is closed under multiplication if and only
if $\Omega$ is a connected subset of $[-1,1]$ of the form $[-\alpha,
\beta],\ 0\leq \alpha\leq 1$ and $\beta \geq \alpha^2$, open boundaries
being also permissible.
\end{thm} 

\begin{proof} ~ Let $(n_1+n_2\t)$, $(m_1+m_2\t ) \in \Sigma(\Omega)$, 
with $\Omega$ as stated 
in the theorem. By definition, $(n_1+n_2 \tau')$, $(m_1+m_2\tau')\in \Omega$. 
Since the multiplication properties of $\t$ and $\t'$ are identical,
the Galois conjugate of this product is the product of the conjugates of 
the factors,  
\be
(n_1+ n_2\tau' )(m_1 +m_2\tau')=((n_1 +n_2\tau)(m_1 +m_2\tau))' \in \Omega.
\ee 
Thus, $(n_1 +n_2\t)(m_1 +m_2\t) \in \Sigma(\Omega)$. 
This proves that the set is closed under multiplication.  \finis 
\end{proof}

In particular, it follows that the Fibonacci chain ${\mathcal F}$ is 
multiplicative\footnote{Specifically, $\forall m_2, n_2 
\in\ \integer$, and 
for $p_2\equiv  m_2n_2 +n_2\left\lfloor\frac{m_2}{\t}+
1\right\rfloor+m_2\left\lfloor\frac{n_2}{\t}  +1\right\rfloor$, 
it follows that $m n = p \in \mathcal{F}$. 
The Fibonacci chain $\mathcal{F}$ is an Abelian monoid.
Another, technical, way to cast this result of the theorem is as an 
identity for the floor function, 
$$ 
\left\lfloor\frac{1}{\tau}\left( m_2 n_2 +m_2 \left\lfloor\frac{n_2 }{\t} 
+1\right\rfloor + 
n_2 \left\lfloor\frac{m_2 }{\t} +1\right\rfloor\right)+1\right\rfloor\,=\,
m_2 n_2 +\left\lfloor\frac{m_2 }{\t} +1\right\rfloor\left\lfloor\frac{n_2 }{\t}
+1\right\rfloor.$$  }. Since $(\t')^{2m} \in (0,1]$, $(\t')^{2m-1} \in [-1,0)$, and $(\t')^m \in [-1,1]$ for $m\in\natural$, the point sets $P_1:=\lbrace \t^{2m}\, \vert \, m \in \natural\rbrace$, $P_2:=\lbrace \t^{2m-1}\, \vert \, m \in \natural\rbrace$ and $P_1\cup P_2$ are examples of multiplicative subsets of ${\mathcal F}$, $-{\mathcal F}$ (i.e. the sequence obtained after reflection at the origin) and $\Sigma([-1,1])$, respectively.

The largest set $\Omega$ for which $\Sigma(\Omega)$ is closed under 
multiplication is $\Omega =[-1,1]$, and we will therefore construct algebras 
over elements in this set. Algebras related to subsets of 
$\Sigma(\Omega)$ that are also closed under multiplication correspond 
to subalgebras of this algebra. 

Due to the property 
\be
\Sigma(\Omega) = -\Sigma(-\Omega)~, \label{minus}
\ee
the set $\Sigma([-1,1])$ decomposes as 
\be 
\Sigma([-1,1]) = -\Sigma((0,1])\oplus \lbrace 0 \rbrace \oplus 
\Sigma((0,1]) , \label{decompose} 
\ee
where $\Sigma((0,1])$ again corresponds to the Fibonacci 
chain ${\mathcal F}$.  

Let $m$, $n\in \Sigma([-1,1])$. Then $mn\in \Sigma([-1,1])$, and hence 
the {\it logarithmic set over $\Sigma([-1,1])$},  defined as 
\be
{\mathcal L}\equiv \lbrace \log(\vert m \vert)~~ \vert ~~ m \in 
\Sigma([-1,1]) \rbrace 
\ee
is additive, with $\log(\vert m \vert) +\log(\vert n \vert) 
=\log(\vert mn \vert)$. Thus the logarithms of the 
set behave additively, and can then index an alternative centerless Virasoro 
algebra on the quasicrystal to that proposed in \cite{patera,twarock}: 
\be 
[L_{\log(\vert m \vert)} ,L_{\log( \vert n \vert)}]\,=\, 
\log \left(\frac{\vert m \vert}{\vert n \vert}\right) 
L_{\log(\vert mn \vert)}~.
\ee 
(One may visualize this through a conventional Witt algebra realization 
$L_{\log(\vert m \vert)}= \vert m \vert^{-x}  \pd_x $.) 

Given the existence of primes on this set, e.g., $5\t+4$, 
the multiplicative composition of indices presents a peculiarity, 
illustrated by the analogous 
multiplicative centerless Virasoro algebra over the integers. Writing the 
$n$th element as $l_n$ (substituting $m\in\integer$ for 
$\log(\vert m \vert)$ etc.), 
\be
[l_m,l_n]\,= \left( m- n \right)l_{mn} . \label{logvir}
\ee
The unusual feature is that if $p$ is prime, then $l_p$ cannot be 
constructed as the commutator of two other elements in the algebra, 
and neither can be $l_p^2$ . 

\begin{rem} 
It is straightforward to extend the definition of cut-and-project 
quasicrystals or model sets to irrational numbers defined as solutions 
$\alpha$ of the quadratic equation $x^2=mx+1$, $m\in \natural$, and  $x^2=mx-1$, $m\in \natural$, $m\geq 3 $,  
(with Galois conjugate $\alpha'$), i.e., to point sets  
\be 
\Sigma_{\alpha}({\Omega}) := \lbrace n\in \integer [\alpha]  ~~\vert ~~ n^*= n_1+ \alpha' n_2\in \Omega \rbrace\,, 
\ee  
where $\integer [\alpha] = \integer + \integer \alpha$. 
Theorem \ref{thm1} holds mutatis mutandis, for $\Omega$ appropriate subsets 
of $[-1,1]$. The sets $\Sigma_{\alpha}([-1,1])$ are also closed under 
multiplication and define algebras along the lines given above. 
\end{rem}

\begin{rem} 
The proof above relies on the set-theoretic properties of $\Sigma(\Omega)$. 
Alternatively, the claim may be proven for the Fibonacci chain ${\mathcal F}$ directly starting from the 
explicit floor function definition and its constrained minimization 
of quadratic integer polynomials. 
\end{rem}

\begin{rem}
In an interesting paper Berman and Moody \cite{BM} introduce a  completely different (non-associative) addition law; $m\ |\!\!- n = \t^2m-\t n$ which has the property that if $m,n\in \Sigma$ so is $\t^2 m-\t n$. Thus this property together with 
the multiplicative law  makes $\Sigma$ into a non-associative ring.
\end{rem}

\section{Additive Grading in $\Sigma([-1,1])$} 
As shown in Theorem \ref{thm1}, the set $\Sigma([-1,1])$ is maximal with 
respect to multiplicative closure, and thus plays a distinguished role. 
Moreover, by virtue of (\ref{decompose}), its properties follow from the 
properties of the Fibonacci chain, i.e., from $\Sigma((0,1])$.
We now further consider graded addition closure of this set.

\begin{lem}\label{lem1}  ~~
Let $n\in \Sigma((0,1])$. If $n\in\Sigma((0,{1\over \t}])$, then  $n+ \tau^2 m \in \Sigma((0,1])$,
while if $n\in\Sigma(({1\over \t} ,1])$, then  $n+ \tau m \in \Sigma((0,1])$.
\end{lem}
\begin{proof}~~
Let $n\in \Sigma((0,1])$. Then, either $n\in \Sigma(({1\over \t} ,1])$, or 
$n\in \Sigma((0,{1\over \t}])$. 
Consider first $n\in \Sigma(({1\over \t} ,1])$; 
by definition, $n^*\in ({1\over \t} ,1]$. Hence, 
$(n + \tau m)^* = n^* + \tau' m^* \in (0,1]$, $\forall m\in \Sigma((0,1])$, 
and thus $n + \tau m \in \Sigma((0,1])$, $\forall m\in \Sigma((0,1])$. 
The claim for $n\in \Sigma((0,{1\over \t}])$ follows analogously. \finis
\end{proof} 

\begin{lem}\label{lem2}
The partition of $\Sigma((0,1])$ in Lemma \ref{lem1} is {\it exclusive}, i.e., 
for every $n\in \Sigma((0,1])$, one has {\it either} 
$n+ \tau m \in \Sigma((0,1])$ {\it or} $n+ \tau^2 m \in \Sigma((0,1])$,
 $\forall m\in \Sigma((0,1])$.  
\end{lem}
\begin{proof}~~ Let $n$, $m\in \Sigma((0,1])$. 
If $n^*+\t' m^*  \not\in (0,1]$, then $n^*$ must lie in $(0,\frac{1}{\t}]$, 
as minimum $n^*+\t' m^* < 0$. But then the above argument shows 
that    $n^*+\t'^2 m^* { \in (0,1]}$. Likewise, if $n^*+\t' m^* \not\in (0,1]$,
 then $n^* \in (\frac{1}{\t},1]$, as maximum $n^*+\t' m^* >1$. \finis
\end{proof}

Due to Lemma \ref{lem2} {\em all} elements $n$ of $\Sigma((0,1])$ may be assigned either an even 
or an odd ``parity", depending on inclusion of their conjugates $n^*$ in the 
respective intervals $(0,1/\t]$ or $(1/\t ,1]$. Lemma \ref{lem1} 
implies that addition of an odd (even) 
element with $\t$ ($\t^2$) times any element of $\Sigma((0,1])$ produces elements of $\Sigma((0,1])$\footnote{
This provides a recursive construction of the chain, starting from, e.g.,
 $m=1$, provided, 
however, that the parities of each new element were at hand. 
Empirically, one may check that the odd class is indexed by 
non-contiguous integers: $n_2= 0,2,5,7,10,...$; while the even class is 
more populous: $n_2= 1,3,4,6,8,9,...$)}. 

\begin{rem}
Depending on the choice of the boundaries of the acceptance window $\Omega$ in (\ref{dual}), the occurrence of an {\it exceptional tile} is possible, i.e. a tile whose nearest neighbour distance (tile) in the chain  occurs precisely once. For example, in the case of the Finonacci chain ${\mathcal F}$, i.e. for $\Sigma((0,1])$, all tiles are of the size $\tau$ or $\tau^2$. However, in $\Sigma([0,1])$, there occurs, in addition, an exceptional tile of size 1, located between the points 0 and 1. \\
The occurrence of exceptional tiles in a point set $\Sigma(\Omega)$ has implications on the assignment of parity to the points. In particular, in the presence of an exceptional tile there occurs precisely one point to which parity cannot be assigned in a unique way: it is that point, which, after removal of one of the boundary points of $\Omega$, lies on the boundaries (open and closed respectively) of the intervals that partition $\Omega$ and determine the parity. 
For example, for $\Sigma([0,1])$, it is the point $n=-\t$ with $n^*=1/\t$. 
\end{rem}

Note, however, that this graded additive composition law is not only noncommutative,
e.g., $(2\t+2) + \t ( 3\t+2) \neq ( 3\t+2) +\t^2  (2\t+2) $. 
It is also  non-associative, e.g. 
\be
(\t+1 + \t^2 (5\t+4 )) + \t(2\t+2) \neq (\t+1)  + 
\t^2 (5\t+4 + \t (2\t+2) ). 
\ee 
Thus, indexing associative algebras over the quasicrystal chain is
incompatible with this grading. 

To recover associativity, one must consider a finer grading 
for the additive composition law of indices. 
Consider, instead, given an integer $a$, those points $n\in \Sigma((0,1])$  
s.t., $\forall m \in \S$, one has  $n+\t^a  m \in \S$. In this case,
we may say that $n$ is {\it compatible with grading $a$}. 
Evidently, $n^* \in (0,1-{1\over \t^a} ]$, for $a$ even; and 
$n^* \in ({1\over \t^a} ,1]$, for $a$ odd. 
Further note that, since the two intervals grow and overlap increasingly with 
growing $a$, a given index point $n$ may be assigned {\em several even and odd} 
gradings $a$, for sufficiently large $a$. This multi-graded 
additive composition law also produces elements in $\Sigma((0,1])$: 

\begin{lem}\label{lem3}~~
For all $n \in \Sigma((0,1])$, compatible with grading $a$ and for all 
$m \in \Sigma((0,1])$, compatible with grading $b$, the point 
$n+\t^a  m$ ( $\in\Sigma((0,1])$ by assumption) is compatible 
with grading $a+b$.  
\end{lem}
\begin{proof} 
There are four cases corresponding to the four possibilities for the gradings 
$a,b$ to be even or odd; in self-evident notation, 
\be
\begin{array}{ll}
(1) &  n^*+\t'^{2r+1}m^* \in [\frac{1}{\t^{2r+1}},1]
-[\frac{1}{\t^{2r+1}}\frac{1}{\t^{2s+1}},1] 
\in (0,1-\frac{1}{\t^{2(r+s+1)}}], \\[.3cm]
(2) & n^*+\t'^{2r+1}m^* \in [\frac{1}{\t^{2r+1}},1]
-\frac{1}{\t^{2r+1}}[0,1-\frac{1}{\t^{2s}}] 
\in (\frac{1}{\t^{2(r+s)+1}},1], \\[.3cm]
(3) &  n^*+\t'^{2r}m^* \in [0,1-\frac{1}{\t^{2r}}]
+\frac{1}{\t^{2r}}[\frac{1}{\t^{2s+1}},1] \in (\frac{1}{\t^{2(r+s)+1}},1],\\[.3cm]
(4) & n^*+\t'^{2r}m^* \in [0,1-\frac{1}{\t^{2r}}]
+\frac{1}{\t^{2r}}[0,1-\frac{1}{\t^{2s}}] \in (0,1-\frac{1}{\t^{2(r+s)}}].
\end{array}
\ee
~~~~~~~~~~~~~~~~~~~~~~~~~~~~~~~~~~~~~~~~~~~~~~~~~~~~~~~~~~~~~~~~~~~~~~~~~~~~~~~~~~~~~~~~~~~\finis
\end{proof}
In contrast to the previous additive composition law, this one 
{\em is} manifestly associative, 
provided the grading of a ``sum" is taken to be the sum of the 
gradings of the ``summands", as here. 

Moreover, the set $\S$ possesses {\it self-similarity} properties: 
\begin{lem} 
Let $a\in \natural$. Then 
\be \label{selfsim}
\tau^a  \Sigma((0,1]) =  \left\lbrace 
\begin{array}{ll}
\Sigma([\frac{-1}{\tau^a},0)) & \mbox{ for } a \mbox{ odd }\\
\Sigma((0,\frac{1}{\tau^a}]) & \mbox{ for } a \mbox{ even }\\
\end{array} \right.\,. 
\ee
\end{lem}
This implies that the sequence of consecutive points in $\Sigma((0,1])$ 
corresponds to that in $\Sigma((0,\frac{1}{\tau^a}])$ (respectively to $\Sigma([\frac{-1}{\tau^a},0))$), up to rescaling of 
all nearest neighbour distances (distances between adjacent points in the set)
by $\tau^a$. 

\begin{proof} ~Let $n\in \tau^a  \Sigma((0,1])$. 
Then, $n= \tau^a m$ with $m^* \in (0,1]$. Hence $n^*=(\tau')^a m^* 
\in (0,(\tau')^a]$ for $a$ even; and $n^* \in [(\tau')^a, 0)$ for $a$ odd. 
This is equivalent to $n\in \Sigma((0,(\tau')^a])$ for $a$ even and 
$n\in \Sigma([(\tau')^a, 0))$ for $a$ odd. 
By the identity $\tau \tau'=-1$, the claim follows. \finis
\end{proof}

Hence, the decomposition (\ref{decompose}) can be refined using  
\be 
\Sigma((0,1]) = (-\tau)  \Sigma((0,1]) 
\oplus \Sigma((\frac{1}{\tau},1]) . \label{decompose2}
\ee
(\ref{decompose2}) again reflects the self-similarity properties of $\Sigma((0,1])$: the set contains a copy of itself in which all distances are rescaled by $\tau$ and the points are reflected at the origin. 
The properties of $\S$ listed in this section have 
implications on the consistent definition and structure of the algebras (\ref{mult}),  as discussed in the next section. 
For example, the selfsimilarity properties lead to a class of subalgebras indexed by a set that is selfsimilar to the original index set.  

\section{Associativity of the algebras}

In order to obtain a consistently associative product (\ref{mult}), 
the finer grading additive composition law of index points 
of Lemma \ref{lem3} is utilized. 
$\forall n \in \Sigma((0,1])$ compatible with integer nonzero grading $a$,
the algebra operators ${J^a_n}$ are defined with 
the additive super-index $a$ as the grading of the graded-additive point $n$ in 
the sub-index.  

The product $J^a_{m}J^b_n\, =\,J^{a+b}_{m+\t^an}$ is thus well defined,
with the respective indices of the right-hand side in their 
proper set. Moreover, the resulting product is manifestly associative, 
as the graded composition of the lower indices is such:
\be
J^a_{n}(J^b_mJ^c_k)\, =\,J^a_nJ^{b+c}_{m+\t^bk}\,
=\,J^{a+b+c}_{n+\t^am+\t^{a+b}k}, \label{mult1}
\ee
whilst the same holds for the alternative association order,
\be
(J^a_{n}J^b_m)J^c_k\, =\,J^{a+b}_{n+\t^am}J^c_k\,
=\,J^{a+b+c}_{n+\t^am+\t^{a+b}k}. \label{mult2} 
\ee 
Associativity follows directly from the compatibility of $n+\t^am$ 
with the additive grading $a+b$ of Lemma \ref{lem3}. 

\begin{rem}
In practice, assigning allowable gradings to a given sub-index is a 
technical problem which has not been studied systematically---beyond the 
case-by-case specific evaluation of interval inclusion of the Galois 
conjugate of the sub-index, or the recursive fine-graded addition of 
starting points compatible with gradings 1 and 2 to build up further 
compatible gradings.  
\end{rem}

With this it is possible to consistently introduce a Lie algebra over the set $\Sigma([-1,1])$ as follows. 
For $J^a_{m}$ with either $m\in \Sigma([-1,0)\cup(0,1])$ and $a\in\natural$, or, $m=0$ and $a=0$,  
\be 
[J^a_{m},J^b_n]\,=\,J^{a+b}_{m+\t^an}\,-\,J^{a+b}_{n+\t^bm}\, 
\ee\label{algebra}
forms a Lie algebra, denoted as ${\mathcal L}([-1,1])$. Since the structure of the commutator is such that the lower central series does not become zero, the Lie algebra is not nilpotent, in contrast to the Lie algebra in (\ref{AW}).

The center of the Lie algebra is given by $J_0^0$. No other element is central since the fine grading $a=0$ is not permissible in combination with $m\not=0$. 

${\mathcal L}([-1,1])$ allows for a triangular decomposition. Indeed, with the notation ${\mathcal L}(\Omega)$ corresponding to the subalgebra with index set $\Sigma(\Omega)$ one has 
\be\label{trianDecomp}
\begin{array}{rcl}
{\mathcal L}([-1,1]) & = & {\mathcal L}([-1,0)) \oplus \lbrace J_0^0 \rbrace \oplus {\mathcal L}((0,1])\\
& = & -{\mathcal L}((0,1]) \oplus \lbrace J_0^0 \rbrace \oplus {\mathcal L}((0,1])\,,
\end{array}
\ee
and the two copies ${\mathcal L}((0,1])$ and ${\mathcal L}([-1,0))$ are related by the automorphism that maps $m\in \Sigma((0,1])$ on $-m \in \Sigma([-1,0))$. This opens up the possibility to study highest weight representations for (\ref{algebra}).  

Independently of this, a simple formal operator realization of this algebra is \cite{FZ}
\be
J^a_m =   e^{m \exp (x)}~ \t^{a \pd_x} . 
\ee

Subalgebras of (\ref{algebra}) are of particular interest because they reflect the selfsimilarity properties of the index set. In particular, by virtue of (\ref{selfsim}), there exists a family of subalgebras, parameterized by $a\in\natural$, 
\be
{\mathcal L}([-{\mid\tau'\mid}^{a},{\mid\tau'\mid}^{a}]) \subset {\mathcal L}([-1,1])
\ee
with index sets selfsimilar to that of ${\mathcal L}([-1,1])$, i.e. with index sets corresponding to copies of $\Sigma([-1,1])$ rescaled by $\tau^{a}$. This property distinguishes the Lie algebras in (\ref{algebra}) from previously defined Lie algebras over aperiodic point sets, and makes them a generic object for the study of quasicrystals. 

\section*{Conclusion}

A closer inspection of the algebras (\ref{algebra}) reveals that they are in fact representatives of a new type of infinite dimensional Lie algebra induced by semi-direct product laws. To see this, note that by virtue of the relation 
\be
\t^{n} = f_n\t +f_{n-1},\ \ \ n>0,\label{example}
\ee 
where $f_n$ denotes the $n$'th Fibonacci number ($f_0 =0,\ f_1=1$), one can express 
$m+\t^{a} n = (m_1+ \t m_2) +  \t^{a} (n_1+ \t n_2)$ as 
$(m_1 +f_an_2 + f_{a-1} n_1) +\t (m_2 +f_a n_1 + f_a n_2 + f_{a-1} n_2)$. In particular, this induces a semi-direct product  
\be\label{semi}
\Sigma(\Omega) \times_\Phi \natural := \lbrace (\widehat{m},a) \, \vert \, \widehat{m}=(m_1,m_2), m=m_1+\tau m_2 \in \Sigma(\Omega), a \in \natural \rbrace 
\ee 
with 
\be 
(\widehat{m},a) \times_\Phi (\widehat{n},b) =(\widehat{m}+\Phi(a) \widehat{n}, a+b)
\ee
where, using the identity $f_{a+1}=f_a +f_{a-1}$,  
\be 
\Phi(a) = \left(
\begin{array}{rcl}
f_{a-1} & f_a \\
f_a & f_{a+1}
\end{array}
\right)\,.
\ee
In this notation, the Lie algebra ${\mathcal L}([-1,1])$ in (\ref{algebra}) can equivalently be expressed as follows\footnote{Note that the Jacobi identity is fulfilled due to the identity $\Phi(a)\Phi(b) = \Phi(a+b)$ for any $a$, $b\in\natural$.}: 
 \be 
[J_{(\widehat{m},a)},J_{(\widehat{n},b)}]\,=\,J_{(\widehat{m},a) \times_\Phi (\widehat{n},b)}\,-\,J_{(\widehat{n},b) \times_\Phi (\widehat{m},a)}\,. 
\ee\label{algebra2} 
It is called the {\it semi-direct product induced} Lie algebra ${\mathcal L}(\Sigma(\Omega) \times_\Phi \natural)$. 

This opens up a new direction in the exploration of infinite dimensional Lie algebras. In particular, Lie algebras as in (\ref{algebra2}) can be associated also to other types of semidirect products, which do not necessarily need to be related to the aperiodic structures $\Sigma(\Omega)$. For example, the cyclotomic case \cite{FZ} may also be viewed from this perspective.   

The semi-direct product induced Lie algebras introduced here for (\ref{semi}) reflect properties of the aperiodic point sets such as their self-similarity. They hence lend themselves for applications in areas where aperiodicity plays a crucial role, such as for example in the study of mathematical and physical properties of quasicrystals \cite{Shechtman:1984}.  They form a Lie algebra of Kac-Moody type, which closes on the points of a one-dimensional
quasicrystal. This immediately raises the prospect of further extensions
of this idea to other aperiodic sets, and also to semi-direct products related to 
the points of higher dimensional aperiodic sets, such as Penrose tilings \cite{Penrose:1974}.
From the point of view of physics it is this last possibility which holds
the most promise of applications to the quantum mechanics of aperiodic
lattices. In particular there is the prospect of constructing a field
theory over the points of an aperiodic chain.

\section*{Acknowledgments}
CKZ was supported by the US Department of Energy, Division of High Energy Physics, Contract W-31-109-ENG-38.
RT gratefully acknowledges financial support via an EPSRC Advanced Research Fellowship.

\end{document}